\documentclass[AMA,STIX2COL]{WileyNJDv5}

\input{preamble.tex}

\begin{document}


\articletype{Research Article}

\received{Date Month Year}
\revised{Date Month Year}
\accepted{Date Month Year}
\journal{Magnetic Resonance in Medicine}
\volume{00}
\copyyear{2025}
\startpage{1}
\raggedbottom

\title{Scan-specific k-space interpolation using deep Koopman autoencoders}

\author[1,2]{Wassim Ben Salah}
\author[4,5]{Jon Cleary}
\author[1,2,3]{Sarah McElroy}
\author[6]{Raphael Tomi-Tricot}
\author[6]{Antoine Naegel}
\author[2]{Sebastien Ourselin}
\author[2]{Jonathan Shapey}
\author[2]{Christos Bergeles}
\author[1,2]{Radhouene Neji}

\authormark{Ben Salah \textsc{et al.}}
\titlemark{Scan-specific k-space interpolation using deep Koopman autoencoders}

\address[1]{%
  \orgdiv{Research Department of Imaging Physics and Engineering},%
  \orgname{ School of Biomedical Engineering and Imaging Sciences, King's College London},%
  \orgaddress{\city{ London}, \country{United Kingdom}}%
}

\address[2]{%
  \orgdiv{Research Department of Surgical \& Interventional Engineering},%
  \orgname{ School of Biomedical Engineering and Imaging Sciences, King's College London},%
  \orgaddress{\city{ London}, \country{United Kingdom}}%
}

\address[3]{%
  \orgdiv{MR Research Collaborations},%
  \orgname{ Siemens Healthcare Limited},%
  \orgaddress{\city{ Camberley}, \country{United Kingdom}}%
}

\address[4]{%
  \orgdiv{Department of Radiology},%
  \orgname{ Guy's and St. Thomas' NHS Foundation Trust},%
  \orgaddress{\city{ London}, \country{United Kingdom}}%
}
\address[5]{%
  \orgdiv{Research Department of Cancer Imaging},%
  \orgname{ School of Biomedical Engineering and Imaging Sciences, King's College London},%
  \orgaddress{\city{ London}, \country{United Kingdom}}%
}
\address[6]{%
  \orgname{ Siemens Healthcare SAS},%
  \orgaddress{\city{ Courbevoie}, \country{France}}%
}

\corres{Wassim Ben Salah, Research Department of Imaging Physics and Engineering. School of Biomedical Engineering and Imaging Sciences, King's College London, London, UK. \email{wassim.ben\_salah@kcl.ac.uk}\\
\textbf{Word count:} 4838 words}


\begin{abstract}
{\textbf{Purpose:} To develop a scan-specific, interpretable k-space interpolation technique for the reconstruction of accelerated MRI data.

\textbf{Methods:} A deep Koopman autoencoder for nonlinear k-space interpolation was developed consisting of an encoder, a linear layer for interpolation and a decoder. The resulting network is implemented using mean square error loss function on k-space neighbourhoods derived from the autocalibration signals (ACS), combining a prediction loss and an autoencoding loss. The approach was compared to GRAPPA and RAKI in 281 retrospectively undersampled brain datasets, using quantitative reconstruction metrics. In a subset of 52 randomly selected images, a clinical reader study compared the proposed approach and RAKI in a forced choice scenario and using a five-point ordinal scoring, assessing sharpness, SNR, aliasing and overall image quality. Additionally, the effect of varying parameters such as the number of autocalibration signals, the contribution of the autoencoding loss, the kernel size, and the latent space dimension for the proposed technique was studied.

\textbf{Results:} The proposed approach showed improved quantitative metrics compared to GRAPPA and RAKI. Visually, the method showed higher resilience to noise amplification and improved preservation of sharp details. In the clinical reader study, the proposed approach was preferred overall to RAKI and yielded higher scores than GRAPPA and RAKI. The proposed method showed robustness to parameter variations.

\textbf{Conclusion:} Deep Learning based k-space interpolation can be achieved using a Koopman autoencoder for k-space neighbourhoods. (228/250)
}
\end{abstract}

\title{Scan-specific k-space interpolation using deep Koopman autoencoders \footnote{Early results from this study were presented at past ISMRM annual meetings \cite{wassimKAE2025, wassimKAE2026}}}
 
\jnlcitation{\cname{%
\author{Ben Salah w.},
\author{Cleary J.},
\author{McElroy S.},
\author{Tomi-Tricot R.},
\author{Naegel A.},
\author{Ourselin S.},
\author{Shapey J.},
\author{Bergeles J.}, and
\author{Neji R.}}%
\ctitle{Scan-specific k-space interpolation using deep Koopman autoencoders} \cjournal{\it Magnetic Resonance in Medicine} \cvol{2025;00(00):1--18}.}
\maketitle

\section{Introduction}

MRI enables non-invasive in-vivo visualisation and assessment of anatomy and function with excellent soft tissue contrast. However, MRI has relatively long acquisition times, which can be a significant limitation in clinical practice. To address this challenge, undersampling strategies combined with tailored reconstruction algorithms \cite{pruessmann1999sense,Griswold2002,lustig2007sparse,breuer2006controlled, block2008suppression} are used to reduce scan time. Parallel imaging \cite{pruessmann1999sense,Griswold2002} is a widely used acceleration technique. Taking multi-channels phased array coils, each receiver channel has a different sensitivity profile, and this multichannel information can be used for reconstruction of undersampled data either in image space (e.g. SENSE \cite{pruessmann1999sense}) or in the k-space domain (e.g. GRAPPA \cite{Griswold2002}). GRAPPA reconstructs unacquired k-space lines by performing a weighted combination of neighbouring lines across all channels, with weights learned from a fully-sampled autocalibration region. Although it is widely employed in clinical imaging, GRAPPA can suffer from a decrease of SNR or aliasing artifacts, especially for high acceleration factors \cite{Huang2006, NANA2008, NANA2010119, ding2015paradoxical}. This is because the estimated GRAPPA coefficients can greatly amplify the noise inherently present in the acquired MRI signals \cite{Huang2006, NANA2008, NANA2010119, ding2015paradoxical} when the GRAPPA inverse problem is ill-conditioned.

Regularisation approaches \cite{lin2006prior,qu2006discrepancy} to the GRAPPA least-square problem have been proposed to reduce noise amplification, but at the cost of introducing aliasing artifacts. Previous studies \cite{Huang2006, NANA2008, NANA2010119,Chang2011} have shown that GRAPPA coefficients exhibit a bias that can become significant with high noise level or high acceleration factors, compared to the derived coefficients if the scan was noise-free, leading to a nonlinear error in the reconstruction \cite{Chang2011}. Nonlinearities are also present due a truncation effect due to the use of a GRAPPA kernel with limited support \cite{lyu2018kernl}.

To reduce the nonlinear errors in GRAPPA, nonlinear GRAPPA techniques \cite{Chang2011,lyu2018kernl,chang2018kernel, wang2019improving} use a kernel method inspired by classical machine learning. The kernel function maps nonlinearly the data into a higher dimensional space for a robust nonlinear estimation. However, choosing the optimal kernel is not trivial and is done in a heuristic manner \cite{Chang2011, samsonov2008optimality,nana2007automatic,qu2005tailored}. In addition, the reconstruction can still present blurring or aliasing artifacts depending on mapping complexity \cite{Chang2011}.

More recently, deep learning methods \cite{Knoll8962951} have gathered interest as they capture complex patterns and can learn representations and nonlinear relationships in a scan-specific manner. For example, RAKI \cite{Akakaya2018} performs a nonlinear estimation of missing k-space lines using a three-layer convolutional neural network (CNN) and autocalibration signals (ACS). Recent work has focused on reducing RAKI's reconstruction time with an optimized GPU implementation \cite{zhang2019optimized} or pre-trained models \cite{pan2025fraki}, and on improving the reconstruction quality with residual networks \cite{zhang2022residual}, iterative reconstruction \cite{dawood2023iterative}, increasing model size and initialising the reconstruction with GRAPPA reconstruction as input (residual estimation found in SPARK \cite{arefeen2022scan}) or using a convolutional recurrent neural network (LORAKI \cite{kim2019loraki}) to propose a deep learning approach to AC-LORAKS \cite{haldar2015autocalibrated} iterative reconstruction. However, in addition to limited interpretability, RAKI suffers from blurring and artifacts due to the network design, since the ReLU activation function can be interpreted as an element-wise multiplication with a mask in k-space, so a convolution in image space \cite{dawoodimage}. Other reconstruction approaches such as LORAKI \cite{kim2019loraki} and SPARK \cite{arefeen2022scan}, also present certain limitations; similar to RAKI \cite{Akakaya2018,zhang2022residual}, LORAKI needs greater number of autocalibration signals, while SPARK k-space correction of artifacts and noise amplification is affected by the chosen initial input reconstruction. Additionally, both techniques have potentially higher computational cost and limited interpretability.

To address these limitations, we propose a scan-specific k-space interpolation technique based on deep Koopman autoencoders \cite{Lusch2018, Brunton2022}. The proposed technique identifies a nonlinear mapping of source k-space neighbourhoods, such that if a linear transformation is applied in the latent space, followed by a decoding operation, the source k-space neighbourhoods are reconstructed faithfully with interpolated missing targets across all coil channels. The proposed network is trained using k-space neighbourhoods extracted from a fully-sampled autocalibration region acquired in the centre of k-space. The deep Koopman autoencoder is by construction interpretable as a manifold interpolation operation, and inherently offers the possibility to control aliasing and blurring artifacts.

\section{Theory}
\label{sec:Deep-Koopman-autoencoder}

\subsection{Deep Koopman autoencoders}

Koopman theory \cite{koopman1931} has been introduced to describe the evolution of nonlinear dynamical systems with the resources developed in classical linear system theory. The Koopman operator \cite{Brunton2022} linearizes nonlinear dynamics for efficient prediction, estimation and control. A short overview of Koopman theory as outlined by Brunton et al. \cite{Brunton2022} is provided below by considering a dynamical system of the form:
\begin{equation}
    \frac{d}{dt}x(t) = f(x(t)), \label{eq:nlPDE}
\end{equation}
where $x \in \chi \subseteq \mathbb{R}^n$ represents the state of the system, contained in a submanifold $\chi$ of a n-dimensional vector space $\mathbb{R}^n$, and $f$ is a vector field describing the dynamics of the system.

A central idea of the theory is to find a global coordinate transformation $z$ to linearize the previous dynamics with an embedding function $\Phi$, with which the dynamics of $z$ are entirely linear and determined by the matrix L:
\begin{equation}
\begin{aligned}
    z(t) &= \Phi(x(t)) \\    
    \frac{d}{dt}z(t) &= Lz(t) \label{eq:lPDE}
\end{aligned}
\end{equation}
The Koopman operator is defined by $\mathcal{K}_{\Phi}\stackrel{\Delta}{=} f\circ \Phi$, and corresponds to the composition of the vector field describing the dynamic $f$ and the embedding function $\Phi$. The Koopman operator is a linear operator that advances the observable $z = \Phi(x)$ and is represented by the continuous-time matrix $L$. Using $F$ and $x_k$ as the temporal discretisation of $f$ the vector field describing the dynamics of the system and of the state system at time $t_{k}=k\Delta t$, respectively, where $\Delta t$ is the discretisation time step and $k$ an integer, the discrete-time nonlinear partial differential equation can be therefore written:
\begin{equation}
\begin{aligned}
    x_{k} &= x(k \Delta t) \\
    x_{k+1} &= F(x_{k}) \label{eq:dnlPDE}
\end{aligned}
\end{equation}
And can be linearized with the discrete-time Koopman operator, with $K$ being the formulation of the continuous-time matrix $L$.
\begin{equation}
\begin{aligned}
    \mathcal{K}_{\Phi}(x_{k}) &= F(z_{k}) = Kz_{k}  \\
    z_{k+1} &= Kz_{k} \label{eq:dlPDE}
\end{aligned}
\end{equation}

Recent research \cite{Brunton2022} demonstrates that Koopman operators can be approximated using data-driven deep learning algorithms such as deep Koopman autoencoders \cite{Lusch2018}. Autoencoders \cite{lecun2015deep} are used to learn meaningful representation or global coordinate transformations. They are traditionally used for dimensionality reduction or feature learning and consist of two parts: an encoder function $h=\Phi(x)$ and a decoder that produces a reconstruction $r=\Psi(h)$ or recovers an input $x=\Psi(h)$. The Koopman operator is learned by an encoder $\Phi$ and a linear layer $K$ from the training data \cite{Lusch2018, GIN2021, Brunton2022, wehmeyer2018time, mardt2018vampnets, takeishi2017learning, yeung2019learning}, the encoder learns a global coordinate transformation, and the linear layer advances the latent observables, while the decoder $\Psi$ recovers the inputs and maps predictions back to the original coordinates. The learning process consists of minimizing a combination of loss functions. A representative example \cite{Lusch2018} is $\mathcal{L} = \mathcal{L}_{\text{rec}} + \lambda_{1}\mathcal{L}_{\text{lin}} + \lambda_{2}\mathcal{L}_{\text{pred}}$, with $\mathcal{L}_{\text{rec}} = \lVert x_{k} - \Psi(\Phi(x_{k})) \rVert$ to learn a useful representation and an accurate recovery of the input, $\mathcal{L}_{\text{lin}} = \lVert \Phi(x_{k+1}) - K\Phi(x_{k})\rVert$ to learn the linear dynamics, $\mathcal{L}_{\text{pred}} = \lVert x_{k+1} - \Psi(K\Phi(x_{k}))\rVert$ to enable future state prediction and encapsulate the overall mapping and estimation process, and $\lambda_1$ and $\lambda_2$ weighting parameters.

\subsection{K-space interpolation described as nonlinear dynamics}

GRAPPA uses a linear convolutional kernel to interpolate unacquired k-space points. The weights are estimated from a fully sampled autocalibration signals ACS as central k-space lines. For each coil $j$ and missing line $m$, a linear least squares fit is solved:
\begin{equation}
    w_{j,m} = \operatorname{arg\,min}_{x} \left\| b - Ax \right\|^2 \label{eq:grappa}
\end{equation}
where $A$ represents the source points in the autocalibration region, b the vector of target points, and $w_{j,m}$ the set of complex weights.

Nonlinear GRAPPA\cite{Chang2011} introduces a kernel method to capture nonlinear relationships, which implicitly defines a nonlinear mapping $\Phi$ applied to the undersampled k-space points, leading to the following minimization problem:

\begin{equation}
    w_{j,m} = \operatorname{arg\,min}_{x} \left\| b - \Phi(A) x \right\|^2 \label{eq:nlgrappa}
\end{equation}
where $\Phi(A)$ is the matrix formed by applying the nonlinear mapping $\Phi$ to each row of the matrix $A$, each row being formed from k-space neighbourhood data across all coils.

An extension to this formulation is to use Koopman autoencoders for nonlinear k-space interpolation. We can learn the nonlinear mapping $\Phi$ with an autoencoder and learn the interpolation in the latent space as a linear dynamic. Therefore, we encode the source points belonging to an undersampled k-space neighbourhood (corresponding to the dynamic system in the deep Koopman autoencoder theory) into a latent space, perform a linear interpolation in this latent space, and then map the result back (decode) to the original space. The input to the network should therefore encapsulate the source k-space neighbourhoods, while the output of the network consists of auto-encoded k-space neighbourhoods (corresponding to the new state of the dynamic system) consisting of the decoded source k-space points and estimated target k-space points.
\section{Methods}

\subsection{Network architecture and implementation}

\begin{figure}[htbp]
    \centering
    \includegraphics[width=0.4\textwidth]{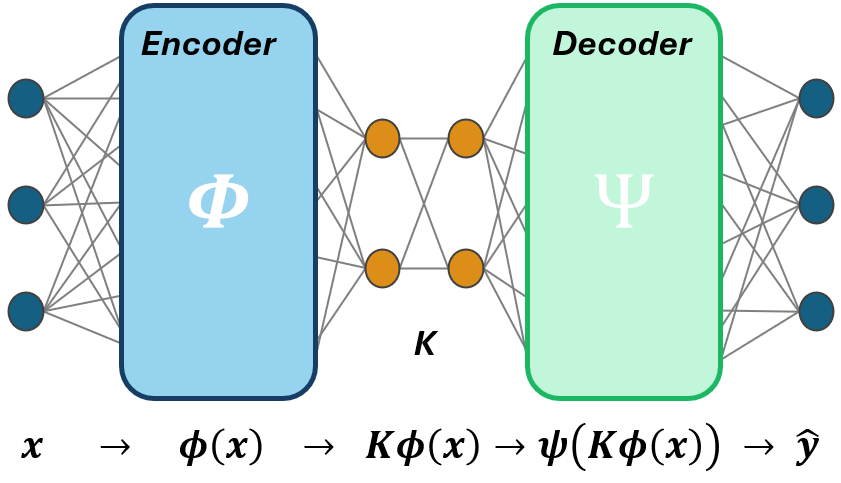}
    \caption[Framework of deep Koopman autoencoder]{The deep Koopman autoencoder consists of a nonlinear mapping via an encoder $\Phi$, followed by a linear interpolation $K$ and a decoder $\Psi$.}
    \label{fig:framework}
\end{figure}

The deep Koopman autoencoder consists of three complex-valued, fully-connected layers: an encoder $\Phi$, a linear interpolation layer $K$ and a decoder $\Psi$. The network architecture is shown in Figure \ref{fig:framework}. The encoder operates on the source points belonging to a k-space neighbourhood, performing a nonlinear mapping to a latent space. The linear layer which encapsulates the system dynamics in standard deep Koopman autoencoders \cite{Lusch2018,GIN2021,Brunton2022} is used in our implementation to perform k-space interpolation in the latent space, and the decoder maps the result back to the original space.
The encoder consists of a bias-free\cite{Akakaya2018, dawood2023iterative} fully-connected layer followed by a ModReLU \cite{arjovsky2016unitary} activation function with a learnable bias parameter $b$ for each node. The nonlinear activation was not used after the bias-free fully-connected decoder layer to not affect the output. The ModReLU \cite{arjovsky2016unitary} is a complex-valued activation, and a variant to ReLU for a complex input $z$:
\begin{equation}
\begin{aligned}
    \operatorname{ModReLU}(z) &= \operatorname{ReLU}(|z| + b) \, \frac{z}{|z|} \\
    &= \begin{cases} 
        (|z| + b) \, \frac{z}{|z|} & \text{if } |z| + b> 0, \\ 
        0 & \text{otherwise,} 
    \end{cases} \label{eq:modrelu}
\end{aligned}
\end{equation}

\begin{figure*}[htbp]
    \centering
    \includegraphics[width=0.9\textwidth]{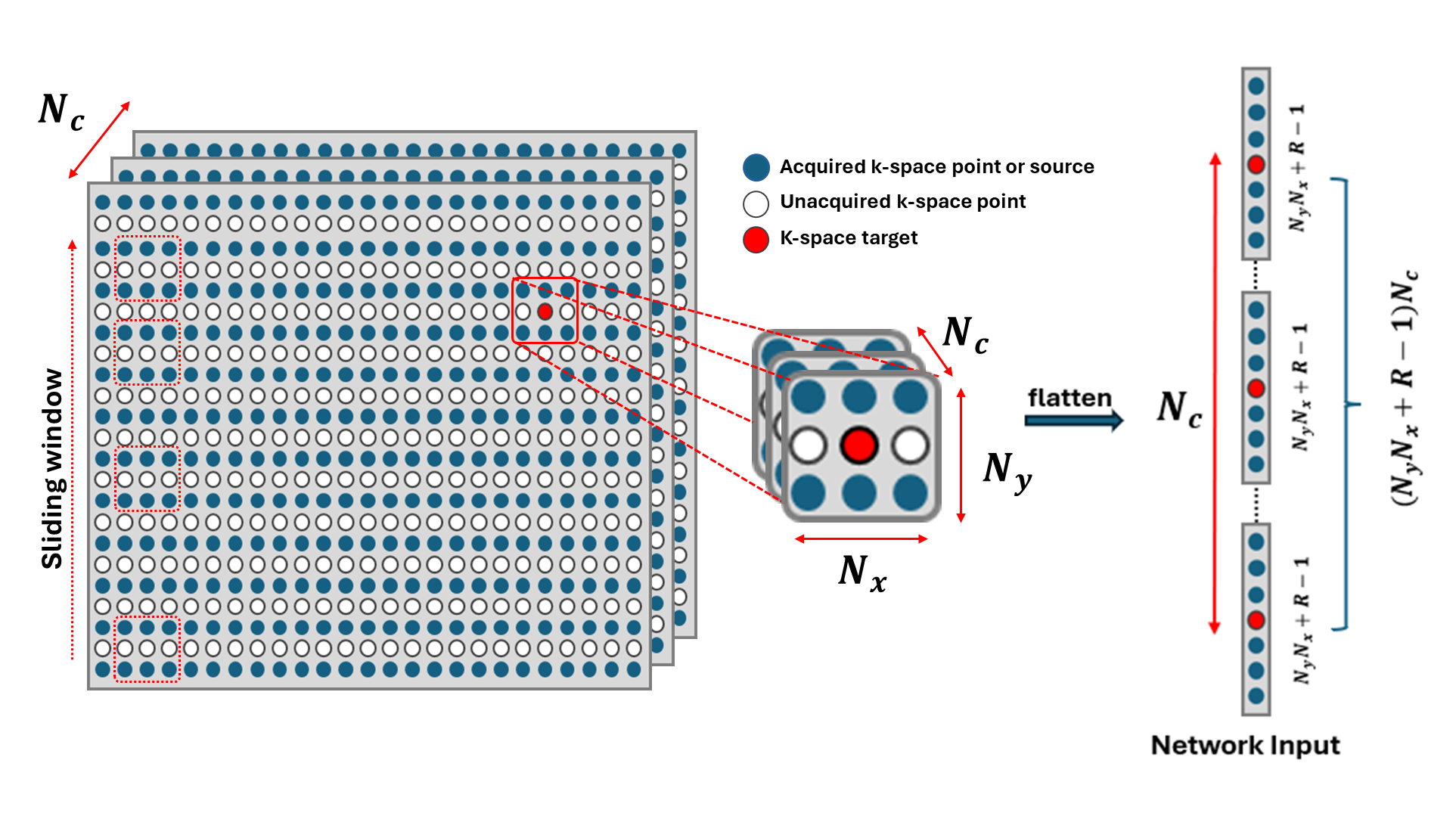} 
    \caption [Extracting input samples for the deep Koopman autoencoder] {Diagram illustrating the input samples used for the deep Koopman autoencoder. For every target position in the undersampled data, a set of sources is extracted across coils and the missing targets are zero-filled. The resulting tensor has a size of \((N_{x} \times N_{y} + R-1) \times N_{c}\), where \([N_{x}, N_{y}]\) is the size of the neighbourhood (kernel size), \(N_{c}\) is the number of coil channels and R is the acceleration factor. The illustration is for \([N_{x}, N_{y}] = [3, 2]\) and R = 2. The operation is applied for every possible position in the autocalibration region to obtain a consequent number of training samples. The obtained tensors of size \((N_{x} \times N_{y} + R-1) \times N_{c}\) are flattened to form the input of the autoencoder architecture. The autoencoder learns to reproduce the input neighbourhoods while interpolating the missing data, and the learning is performed end-to-end.}
    \label{fig:formatting}
\end{figure*}

Let $N_c$ be the number of channels, $R$ be the acceleration factor, $N_x$, $N_y$ the kernel dimensions along the frequency and phase encoding directions, respectively. The input $\bm{x}$ to the network will contain a zero-filled $k$-space neighbourhood, centred at a $k$-space location $\bm{k} = [k_{x}, k_{y}]^T$. The vector $\bm{x}$ is therefore formed by concatenating the multichannel $k$-space samples from the undersampled $k$-space within a local kernel (Figure~\ref{fig:formatting}):

\begin{equation}
    \bm{x} = \begin{bmatrix} \bm{S}^{\text{zf}}_1 & \bm{S}^{\text{zf}}_2 & \dots & \bm{S}^{\text{zf}}_{N_c} \end{bmatrix}^T \in \mathbb{C}^{N_c(N_x N_y + R - 1)},
\end{equation}

$\bm{S}^{\text{zf}}_l$ denotes the zero-filled neighbourhoods of the $l$-th coil of size $N_x N_y + R - 1$. This neighbourhood contains the sources ${S_l(k_{x} - b_x \Delta k_x, k_{y} - R b_y \Delta k_y)}_{\substack{|b_x| \le \lfloor N_x/2 \rfloor \\ |b_y| < \lfloor N_y/2 \rfloor}}$, while the target (unsampled) locations are set to $0$. Similarly, the corresponding output $\bm{y}$ of the network is:

\begin{equation}
    \bm{y} = \begin{bmatrix} \bm{S}^{\text{full}}_1 & \bm{S}^{\text{full}}_2 & \dots & \bm{S}^{\text{full}}_{N_c} \end{bmatrix}^T \in \mathbb{C}^{N_c(N_x N_y + R - 1)},
\end{equation}

$\bm{S}^{\text{full}}_l$ represents the corresponding fully sampled neighbourhood. The deep Koopman autoencoder aims at reproducing the input source data while interpolating the (missing) target k-space data across all coil channels.

To train the network, training pairs $(\bm{x}, \bm{y})$ are constructed by sliding the kernel across all possible calibration locations $\bm{k}$ in the $k$-space ACS region. The network first encodes the zero-filled subsampled k-space neighbourhoods across channels $\bm{x}$, then interpolates the resulting latent vector $\Phi(\bm{x})$ using the linear layer $K$, and finally decodes the reconstructed latent vector $K\Phi(\bm{x})$. The predicted vector $\mathbf{\hat{y}} = \Psi(K\Phi(\bm{x}))$ is compared to the ground truth vector $\bm{y}$. The model is trained with all training data at once (batch gradient descent \cite{lecun2015deep}) using the squared error loss and Adam optimizer \cite{sutskever2013importance,kingma2014adam} with a learning rate of $10^{-3}$ over 4000 epochs. The loss is defined as:
\begin{equation}
    \mathcal{L} = \sum_{i=1}^{N} \left\lVert \bm{y}_i - \bm{\hat{y}}_i \right\rVert_2^{\,2} = \sum_{i=1}^{N} \left\lVert \bm{y}_i - \Psi(K\Phi(\bm{x}_i)) \right\rVert_2^{\,2}
\end{equation}
Where $(\bm{x}_i, \bm{y}_i)$ are training pairs extracted from the ACS, and N is the total number of extracted neighbourhoods. This loss can be rewritten as $\mathcal{L} = \mathcal{L}_{\text{pred}}+ \mathcal{L}_{\text{ae}}$, with:
\begin{equation}
\begin{aligned}
    \mathcal{L}_{\text{pred}} &= \sum_{i=1}^{N} \left\| M_{t} \odot (\bm{y}_i - \bm{\hat{y}}_i) \right\|_2^2 \\
    \text{and} \quad \mathcal{L}_{\text{ae}} &= \sum_{i=1}^{N} \left\| M_{s} \odot (\bm{y}_i - \bm{\hat{y}}_i) \right\|_2^2
\end{aligned}
\end{equation}
and where $\odot$ is the element-wise multiplication, $M_{s}$ and $M_{t}$ are the binary masks for the sources and for the targets, respectively. The elements of $M_{s}$ are equal to 1 at the source locations and zero otherwise, while the elements of $M_{t}$ are equal to 1 at the target locations and zero otherwise. Consequently, the deep Koopman autoencoder implicitly minimizes a combination of an autoencoding loss $\mathcal{L}_{\text{ae}}$ and a prediction loss $\mathcal{L}_{\text{pred}}$ with an equal weighting. Similarly to the loss function used for deep Koopman autoencoders, as introduced in section \ref{sec:Deep-Koopman-autoencoder}, we generalize this formulation by introducing a weighting parameter $\lambda$ to control the contribution of the autoencoding loss term to the total loss:

\begin{equation}
        \mathcal{L} = \mathcal{L}_{\text{pred}}+ \lambda \cdot \mathcal{L}_{\text{ae}}
\end{equation}

The encoder layer was initialised using He normal initialisation \cite{he2015delving} to account for the ModReLU activation function, while the linear interpolation layer and decoder layer were initialised using Glorot normal initialisation \cite{glorot2010understanding}. The weights of each complex-valued fully connected layer were initialised such that the real and imaginary parts are independently sampled from a normal distribution, with the variance scaled according to the nonlinearity applied (via a gain factor) as well as the input and output sizes of the layer \cite{trabelsi2017deep}. The biases introduced by the ModReLU activation functions were initialised to a null vector and were learned during the training. The deep Koopman autoencoder was implemented in PyTorch v2.9.1. The source code is made publicly available\footnote{\url{https://github.com/bensalah-wassim/koopman-autoencoder-mri-reconstruction}} for research purposes.

\subsection{Experiments}

In this study, unless stated otherwise, a kernel size of [5, 4] was used to match the kernel size used in the previously published RAKI study \cite{Akakaya2018} and the output dimension of the encoder, representing the dimension of the latent space, is set to 80\% of the input dimension.
All experiments were performed on a desktop computer equipped with an Intel(R) Xeon(R) w3-2435 (8 cores, 3.10 GHz), 64 GB RAM, and an NVIDIA RTX A2000 with 12GB VRAM, running Windows 11.

\subsubsection{Datasets}
This study used 281 fully sampled raw k-space brain MRIs from the NYU fastMRI dataset \cite{zbontar2018fastmri, knoll2020fastmri} acquired on 1.5T and 3T magnets using multichannel head coils. The raw k-space data included 79 axial T1-weighted (including pre- and post-contrast data), 175 T2-weighted and 27 FLAIR images. The number of coil channels varied between 16 and 20 channels depending on the used head coil. Ten slices for each fully sampled raw k-space volume were used. In all experiments, the raw fully sampled k-space data were retrospectively undersampled using uniform Cartesian undersampling with an acceleration factor of 4. A central k-space region consisting of 40 k-space lines was kept to provide autocalibration signals. The k-space data were normalized so that the maximum k-space magnitude was equal to 1 across all slices for each dataset. The regularization parameter $\lambda$ weighting the contribution of the autoencoding loss was set to 0.15, 0.16, 0.09 and 0.12 for pre-contrast T1-weighted images, post-contrast T1-weighted images, T2-weighted images and FLAIR images, respectively. The regularisation parameter was selected visually for each image contrast to provide the best balance between noise suppression and preservation of anatomical detail.

\subsubsection{Image reconstruction assessment using quantitative metrics}
\label{subsubsec:quantmetrics}

Image reconstruction quality of the proposed approach was assessed in comparison to the baseline methods GRAPPA \cite{Griswold2002} and RAKI (implementation provided by the authors \cite{Akakaya2018} originally in TensorFlow and adapted to PyTorch). For all approaches, sum-of-squares coil combination was performed to obtain the final image. The obtained images were compared qualitatively through visual inspection of the reconstructed images and their absolute differences to the reconstruction of the fully sampled data, as well as using quantitative metrics including root mean squared error (NRMSE) \cite{wang2004image}, structural similarity index (SSIM) \cite{sara2019image, wang2004image} and peak-signal noise ratio (PSNR) \cite{wang2004image}. As suggested in previous work \cite{dawoodimage, nimje2025insights}, the metrics were computed over a brain mask derived from the ground truth images, using the Triangle thresholding method \cite{zack1977automatic}.

\subsubsection{Clinical reader study}
A neuroradiologist (>7 years of experience) reviewed 52 randomly selected volumes (13 volumes for each of the 4 contrasts, out of 281 in total) for the proposed approach and RAKI in two different scenarios: first, a 5-point ordinal evaluation (1 = non-diagnostic, 5 =excellent) assessing perceived sharpness, perceived signal-to-noise ratio (pSNR), presence of aliasing artefacts, and overall image quality; and second, a forced-choice preference assessment comparing two methods presented side-by-side in a randomized order. Statistical testing was performed using a Wilcoxon signed-rank test and a two-sided McNemar test for the first and second scenario, respectively. A p-value < 0.05 indicated statistical significance.

\subsubsection{Assessment of the effects of parameter variation}
For the following experiments, a subset of four non-adjacent slices for each of the 52 randomly selected volumes were selected to reduce processing time while still capturing a range of inter-subject and intra-subject variability. Each of the parameter variations was assessed both visually and using the previously mentioned quantitative metrics (subsection \ref{subsubsec:quantmetrics}).

\vspace{0.5em}
\noindent\textbf{Weighting of the autoencoding loss}\\
The effect of varying the weight of the autoencoding loss on the resulting reconstruction from the deep Koopman autoencoder for the 13 FLAIR images was assessed. The weight $\lambda$ was varied from 0.0 (where the training is conducted based on the reconstruction of missing k-space points, similarly to RAKI) to 1.0, the following values were tested: [0.0, 0.05, 0.1, 0.12, 0.2, 0.4, 0.7, 1.0].

\vspace{0.5em}
\noindent\textbf{Number of autocalibration lines}\\
The dependency of the reconstruction quality on the number of autocalibration lines provided for training/model estimation was evaluated for all approaches. The following numbers of autocalibration lines were evaluated: 24, 30 and 40 lines.

\vspace{0.5em}
\noindent\textbf{Kernel size}\\
The effect of the kernel size on the reconstruction results of the proposed approach and GRAPPA was investigated. The following kernel sizes \([k_x, k_y]\)  were tested: [5, 4], [5,2], [5,6], [3, 4], [5, 4], [7,4], and [9,4].

\vspace{0.5em}
\noindent\textbf{Latent space dimension}\\
The effect of changing the latent space dimension in the deep Koopman autoencoder on the ability of the model to accurately reconstruct the image was examined based on visual assessment and quantitative metrics. Both an overcomplete configuration (latent space dimension higher than input space dimension) and an undercomplete configuration (latent space dimension lower than input space dimension) of the autoencoder were tested, with the ratio between the latent and input dimension varied from 0.2 to 2, with an increment of 0.2.

\section{Results}

\subsection{Quantitative reconstruction quality metrics}

\begin{figure}[tbp]
    \centering 
    \includegraphics[width=0.45\textwidth]{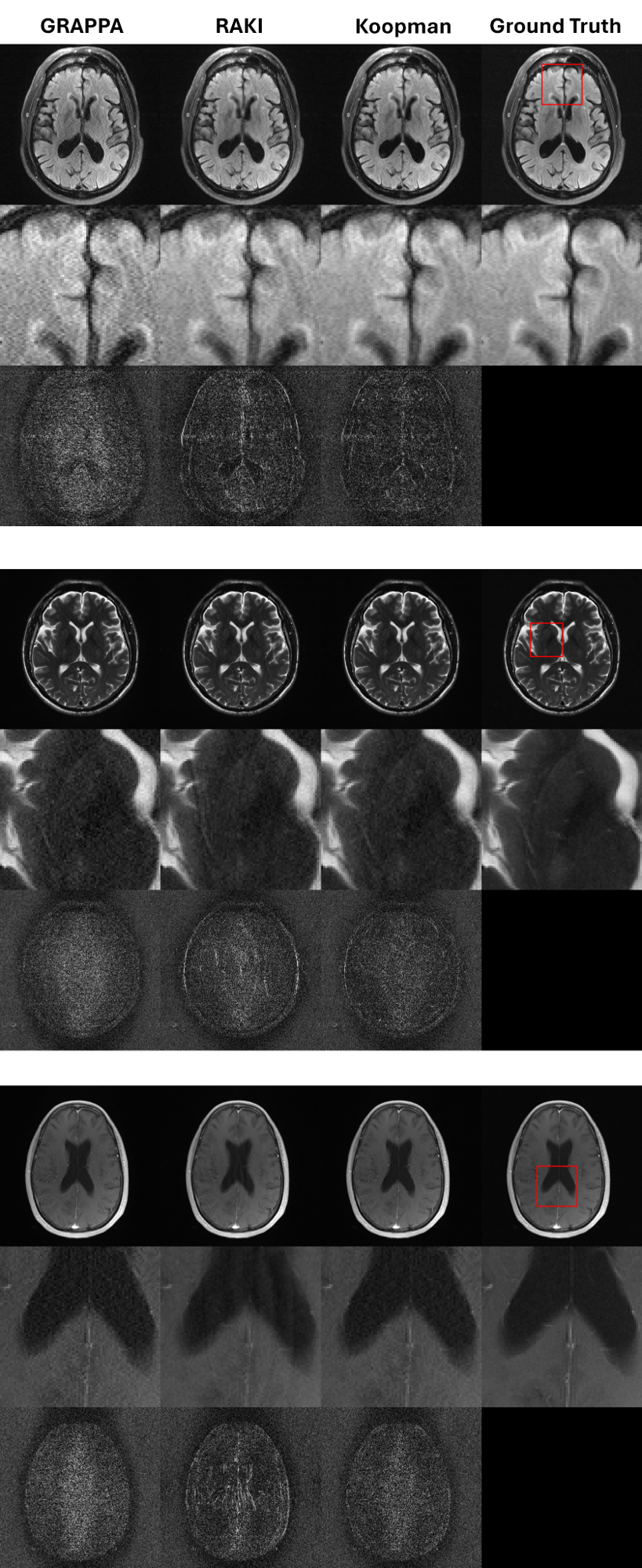} 
    \caption[Comparison of reconstruction performance on the FastMRI dataset at acceleration factor $R=4$]{Comparison of reconstruction performance at acceleration factor $R=4$ for three different contrasts (FLAIR images in top section, T2-weighted images in middle section and T1-weighted images in bottom section). For each method, a reconstructed image is shown (top row), as well as a zoomed-in region highlighted in the full image with a red rectangle (middle row). The bottom row highlights the absolute difference between the ground truth and the reconstructed images.}
    \label{fig:r1_merged}
\end{figure}

Table \ref{tab:acc4_ACS40_metrics} summarizes the quantitative reconstruction metrics for GRAPPA, RAKI and the proposed approach. The proposed approach showed improved SSIM, PSNR and NRMSE compared to both GRAPPA and RAKI, with RAKI yielding improved quantitative metrics compared to GRAPPA. Figure \ref{fig:r1_merged} shows three different representative cases. Visual inspection of the reconstructed images and of the absolute differences to the ground truth images shows higher resilience to noise amplification and sharper details with the proposed approach compared to GRAPPA and RAKI. The absolute difference images highlight reduced residual aliasing artifacts for the proposed techniques compared to RAKI. Additionally, Table \ref{tab:acc4_ACS40_metrics} shows the reconstruction times for the proposed approach and for RAKI. The proposed approach is slightly faster than RAKI (mean reconstruction times: 171s and 212s, respectively).

\begin{table}[htbp]
    \centering
    \begin{tabular}{lrrrrr}
        \toprule
        \textbf{Model} & \textbf{Mean} & \textbf{Std} & \textbf{Median} & \textbf{Q1} & \textbf{Q3} \\
        \midrule
        \multicolumn{6}{c}{\textbf{SSIM $\uparrow$}} \\
        GRAPPA      & 0.8179 & 0.0951 & 0.8429 & 0.7828 & 0.8791 \\
        Koopman     & \textbf{0.8870} & \textbf{0.0515} & \textbf{0.9022} & \textbf{0.8721} & \textbf{0.9180} \\
        RAKI        & 0.8784 & 0.0586 & 0.8960 & 0.8615 & 0.9140 \\
        \midrule
        \multicolumn{6}{c}{\textbf{PSNR $\uparrow$}} \\
        GRAPPA      & 32.5544 & 3.4828 & 33.6473 & 31.1614 & 34.7313 \\
        Koopman     & \textbf{34.8} & \textbf{2.9789} & \textbf{35.4683} & \textbf{33.6975} & \textbf{36.5969} \\
        RAKI        & 34.4084 & 3.0211 & 35.0629 & 33.0913 & 36.2627 \\
        \midrule
        \multicolumn{6}{c}{\textbf{NRMSE $\downarrow$}} \\
        GRAPPA      & 0.0740 & 0.0270 & 0.0679 & 0.0587 & 0.0800 \\
        Koopman         & \textbf{0.0557} & \textbf{0.0136} & \textbf{0.0531} & \textbf{0.0475} & \textbf{0.0605} \\
        RAKI        & 0.0585 & 0.0148 & 0.0557 & 0.0486 & 0.0653 \\
        \midrule
        \multicolumn{6}{c}{\textbf{Reconstruction time (s) $\downarrow$}} \\
        Koopman     & \textbf{171} & \textbf{37} & \textbf{162} & \textbf{136} & \textbf{197} \\
        RAKI        & 212 & 50 & 214 & 181 & 234 \\
        \bottomrule
    \end{tabular}
    \caption{Reconstruction quality metrics (SSIM, NRMSE, PSNR) and reconstruction times for the different reconstruction techniques over 281 brain MRI datasets with acceleration factor 4 and 40 autocalibration lines. Reported: mean, std, median, Q1 and Q3.}
    \label{tab:acc4_ACS40_metrics}
\end{table}

\subsection{Clinical reader study}
\begin{figure}[tbp]
    \centering
    \includegraphics[width=0.45\textwidth]{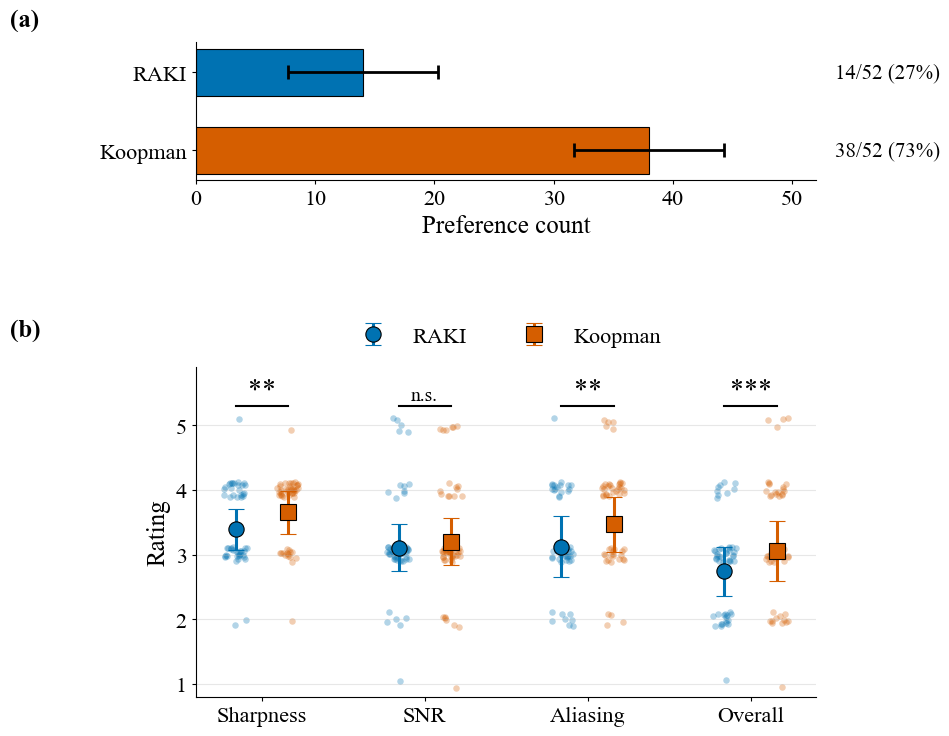}
    \caption[Reader study statistical tests]{Distribution of the preference test results between RAKI and deep Koopman autoencoder with 95\% confidence intervals (a), and the ordinal assessment for the different rating criteria for both techniques (b), Markers indicate mean ratings with 95\% confidence intervals; dots show individual ratings (jittered). Paired Wilcoxon tests: $^{*}p<0.05$, $^{**}p<0.01$, $^{***}p<0.001$, n.s.: not significant.}
    \label{fig:rder_st}
\end{figure}

The reader study results are summarized in Figure \ref{fig:rder_st}. The ordinal qualitative assessment showed that the deep Koopman autoencoder provided a statistically significant improvement compared to RAKI in terms of sharpness (p = 0.0017), aliasing artefacts (p = 0.0028), and overall image quality (p = 0.0003), while the improvement in SNR was not statistically significant (p = 0.3458). In the forced-choice scenario, the proposed approach was preferred to RAKI in 73\% of the considered cases (38 out of the 52 cases), and this preference was statistically significant (p= 0.0003).

\subsection{Effects of parameter variation}

\noindent\textbf{Weighting of the autoencoding loss}\\

\begin{figure*}[tbp]
    \centering
    \includegraphics[width=0.8\textwidth]{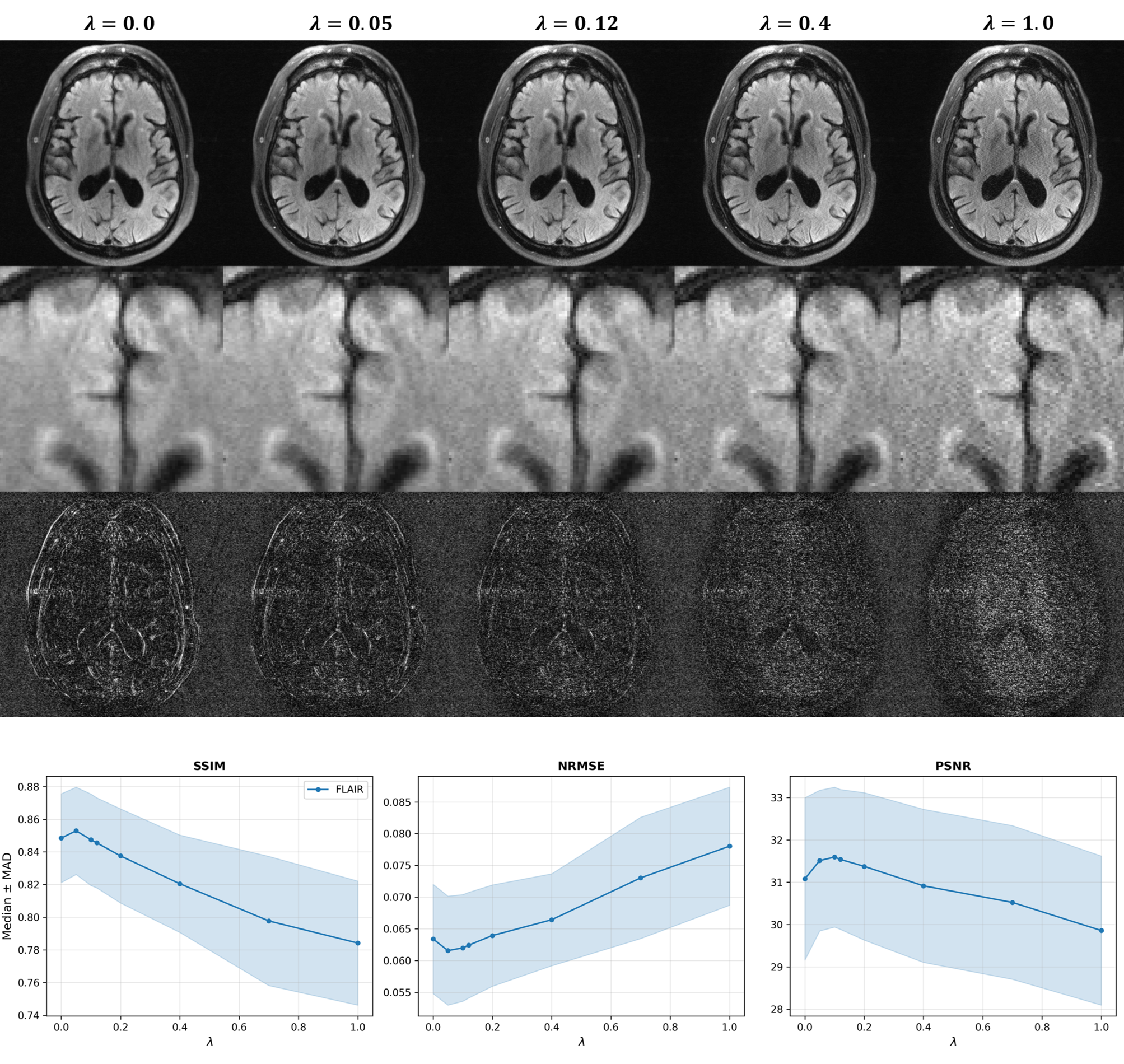}
    \caption[Deep Koopman autoencoder reconstruction using ]{
    Deep Koopman autoencoder reconstruction using different values of the weight $\lambda$, controlling the contribution of the regularization made by the autoencoding loss. For each tested value of the parameter $\lambda$, a reconstructed slice is shown (first row) as well as a zoomed-in view of the reconstruction results (second row). The third row shows the absolute difference between the ground truth and the reconstructed image. Additional plots (fourth row) show the obtained quantitative metrics SSIM, NRMSE and PSNR with respect to the value of $\lambda$ for a subset of 13 FLAIR images. The solid line presents the mean value of each metric, and the shaded error bands represent the standard deviation.
    }
    \label{fig:lambda_fig}
\end{figure*}

Figure \ref{fig:lambda_fig} illustrates the visual and quantitative evolution of image reconstruction for the proposed approach across the tested range of the weight $\lambda$ of the autoencoding loss for the selected subset of FLAIR images. Model performance decreases when $\lambda$ approaches zero, leading to severe under-regularisation and resulting in a degradation of reconstructed image characterized by aliased and over-smoothed regions. 
Conversely, choosing a too high value for $\lambda$ leads to a high level of noise amplification. Intermediate values of $\lambda$ (around 0.12) represent a good trade-off between an over-smoothed image and a noise-dominated image. The obtained quantitative metrics exhibit a similar trend as shown in Figure \ref{fig:lambda_fig}.\\

\noindent\textbf{Number of autocalibration lines}\\

\begin{figure}[tbp]
    \centering
    \includegraphics[width=0.46\textwidth]{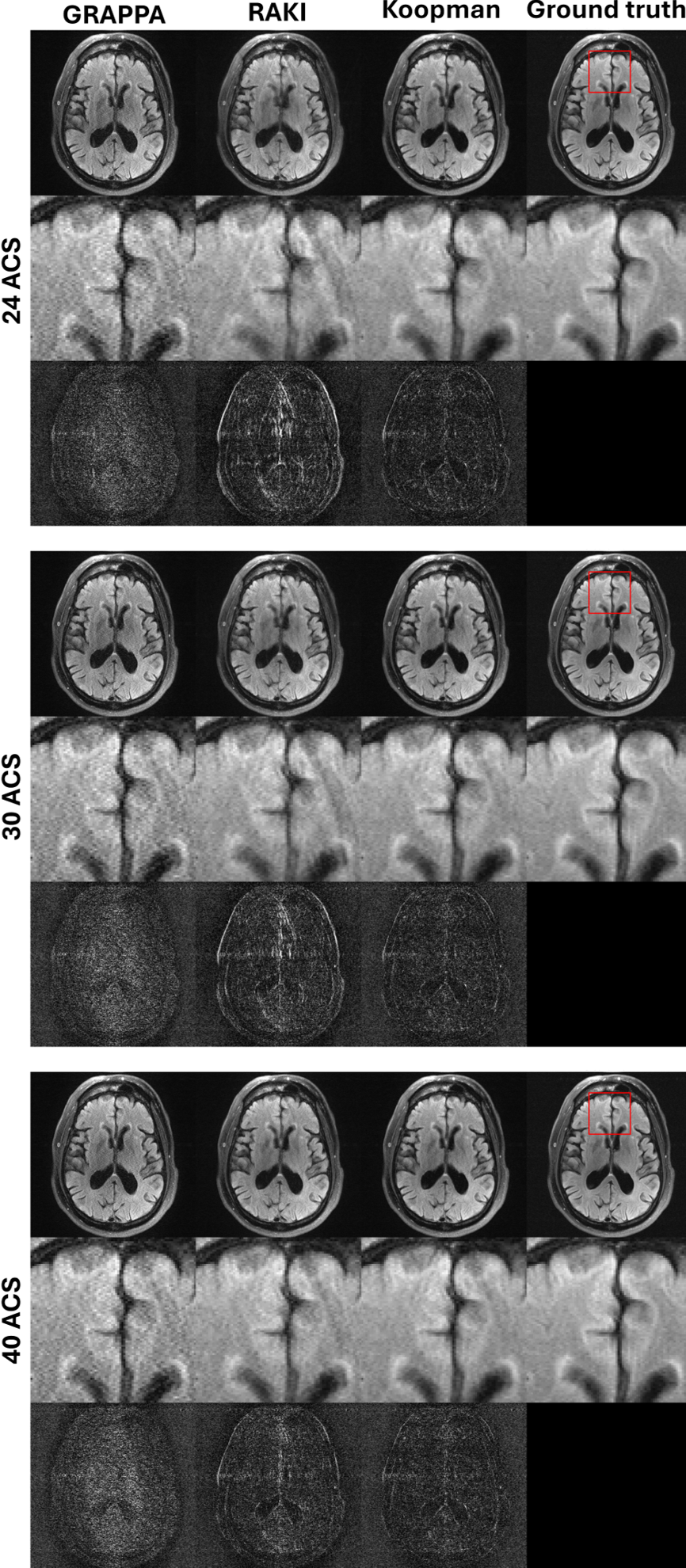}
    \caption[Comparison of reconstruction performance for the different approaches with different number of ACS lines at acceleration $R=4$]{
    Comparison of reconstruction results for the different approaches with different number of ACS lines at acceleration $R=4$ for a representative case. The reconstruction for each method is shown successively for 24, 30 and 40 autocalibration lines. For each number of autocalibration lines, reconstructions of a FLAIR image for different ACS lines are shown (top rows), as well as a zoomed-in region highlighted in the full image with a red rectangle (middle rows). The bottom rows highlight the absolute difference between the ground truth and the reconstructed image for the different approaches.} 
    \label{fig:ACSmerged}
\end{figure}

\begin{figure*}[tbp]
    \centering
    \includegraphics[width=1.0\textwidth]{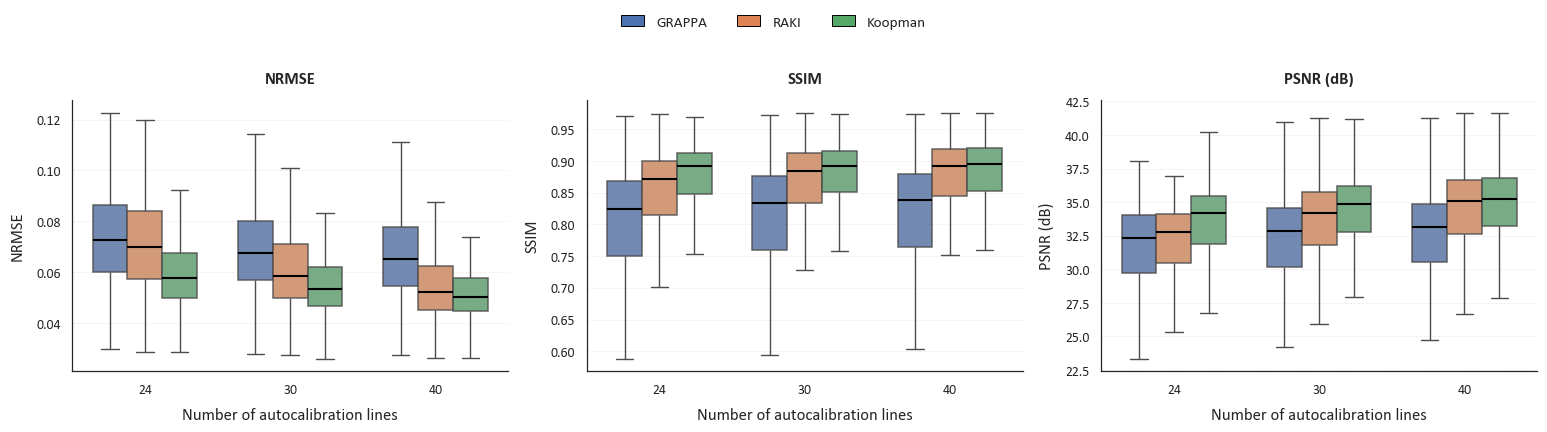}
    \caption[Box plot comparison of the SSIM, NRMSE and PSNR reconstruction metrics for GRAPPA, RAKI and the proposed approach using different numbers of autocalibration lines at an acceleration factor of $R=4$]{Box plot comparison of the SSIM, NRMSE and PSNR reconstruction metrics for GRAPPA, RAKI and the proposed approach using different numbers of autocalibration lines at an acceleration factor of $R=4$. Each subplot highlights the performance of each reconstruction using 24, 30 and 40 autocalibration lines, respectively. The central line in each box represents the median, and the box displays the interquartile range (IQR).} 
    \label{fig:rACS}
\end{figure*}

Figure \ref{fig:ACSmerged} shows a representative example of the reconstructions of GRAPPA, RAKI and the proposed approach, obtained using different numbers of autocalibration lines. With 24 calibration lines, severe noise amplification in the GRAPPA reconstruction and pronounced artifacts in the reconstruction of RAKI can be observed, along with higher noise levels compared to the proposed approach. As more autocalibration lines are provided, the reconstruction results of GRAPPA and RAKI are less noisy and RAKI exhibit fewer artifacts. The reconstruction results provided by the proposed approach remain sharper and more resilient to noise amplification and artifact presence across the different numbers of autocalibration lines. These findings are corroborated by the quantitative metrics shown in Figure \ref{fig:rACS}. With 24 autocalibration lines, the proposed approach shows better NRMSE, SSIM and PSNR metrics than RAKI and GRAPPA. When the number of autocalibration lines is increased, it can be observed that the NRMSE and SSIM metrics of RAKI and GRAPPA get closer to those of the proposed approach, while staying lower.\\

\noindent\textbf{Kernel size}\\

\begin{figure*}[b]
    \centering
    \includegraphics[width=0.95\textwidth]{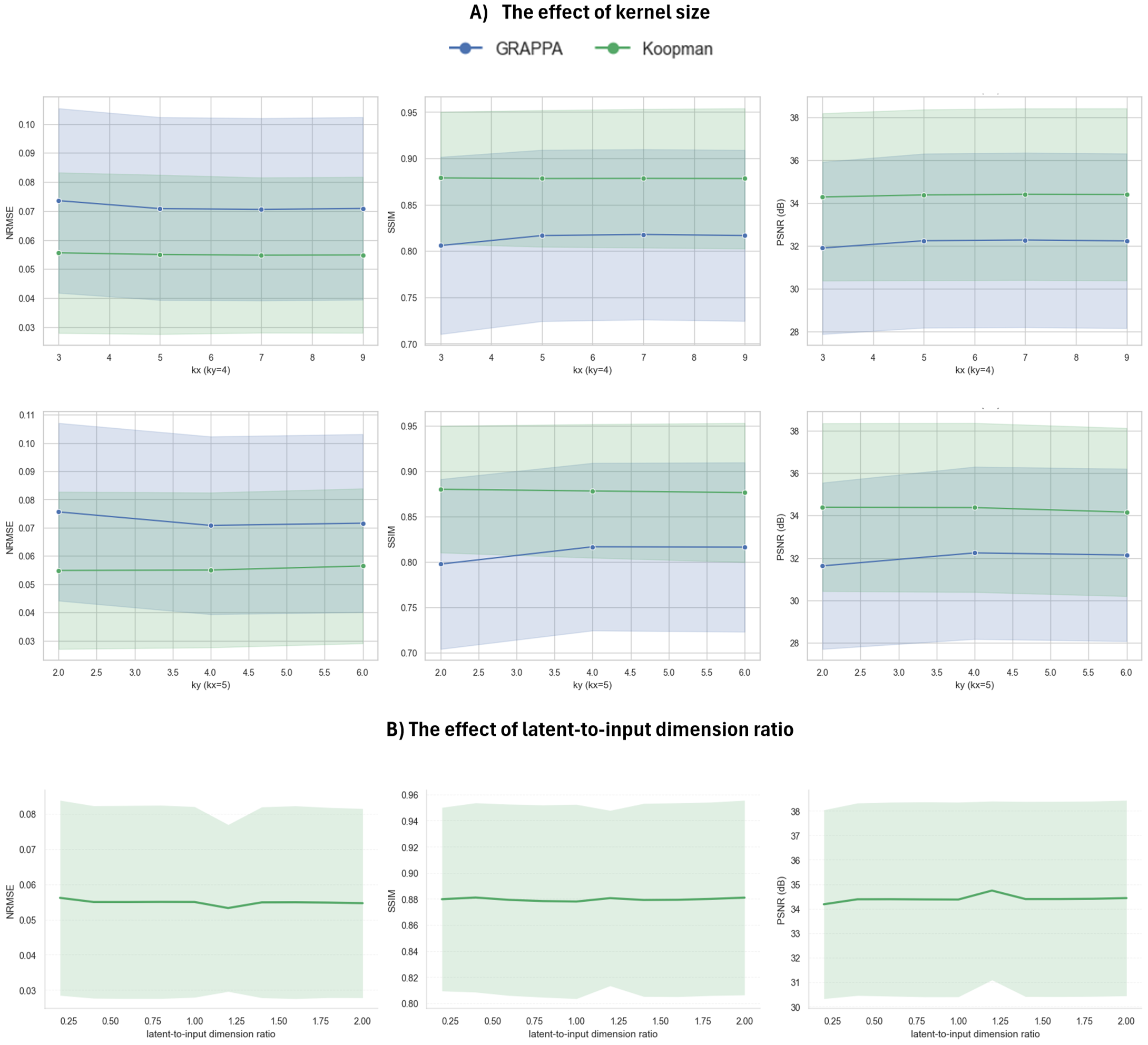}
    \caption[Effect of latent-to-input dimension ratio on deep Koopman autoencoder Reconstruction performance, and effect of Kernel geometry on GRAPPA and deep Koopman autoencoder reconstruction]{
        Section A shows the effect of kernel size on GRAPPA and deep Koopman autoencoder reconstruction using SSIM, PSNR and NRMSE metrics. The top row of section A shows the effect of changing the kernel size in the frequency encoding direction ($k_x = 3, 5, 7, 9$) with a fixed kernel size in the phase encoding direction ($k_y=4$). The bottom row of section A shows the effect of changing the kernel size in the phase encoding direction ($k_y=2, 4, 6$) with a fixed kernel size in the frequency encoding direction ($k_x = 5$). Section B shows the effect of latent-to-input dimension ratio on deep Koopman autoencoder reconstruction performance. The plots present the mean value of each metric over 281 brain MRI datasets as a solid line, with shaded error bands representing the standard deviation.} 
    \label{fig:error_dim} 
\end{figure*}

\begin{figure*}[tbp]
    \centering
    \includegraphics[width=0.8\textwidth]{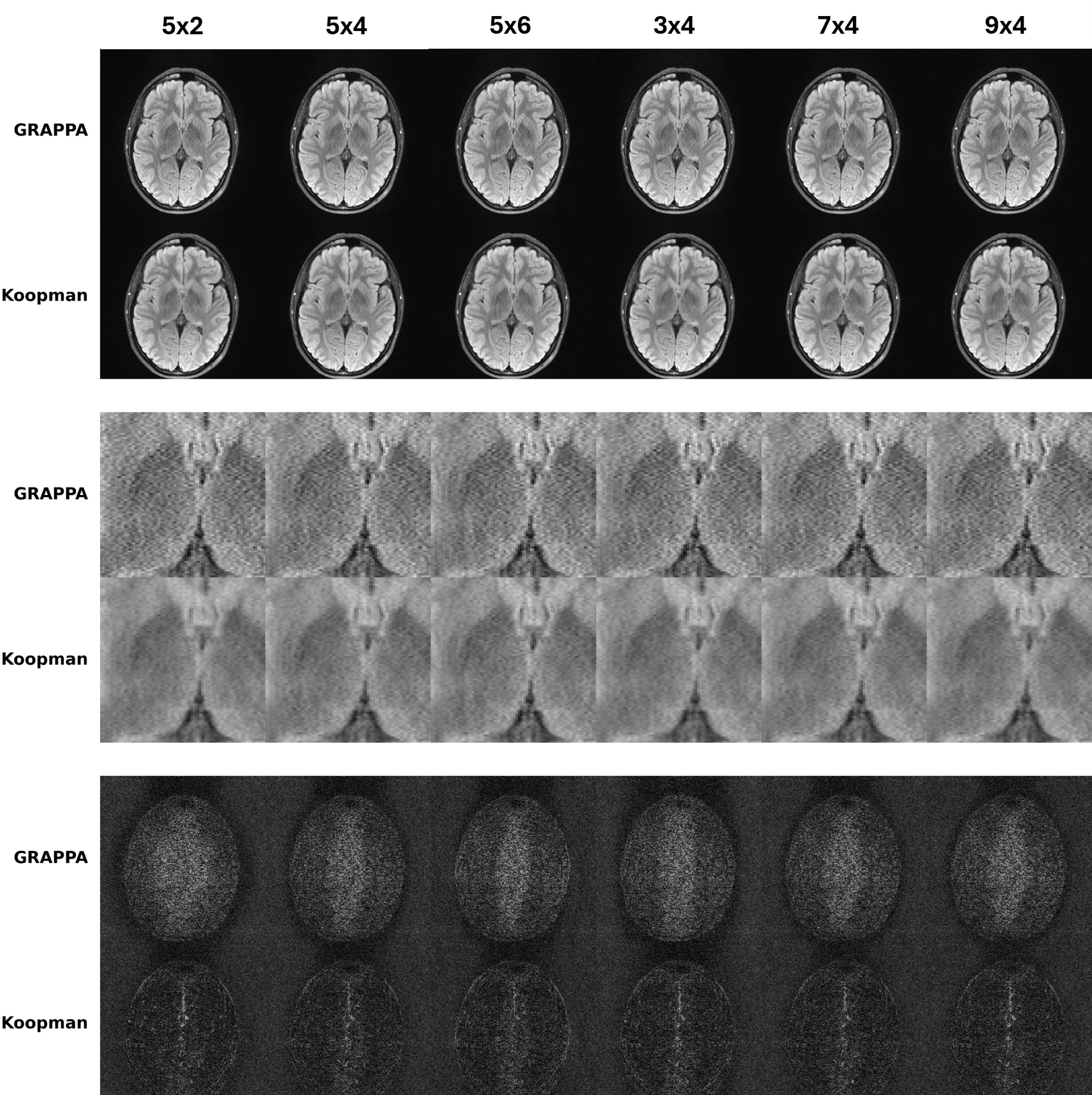} 
    \caption[GRAPPA and deep Koopman autoencoder reconstruction with different kernel size]{Comparison of GRAPPA and deep Koopman autoencoder reconstructions using a kernel size of [5, 2], [3, 4], [5, 4], [7, 4], [5, 6] and [9, 4]. For each kernel size, the reconstruction of a brain image is shown (first two rows), as well as a zoomed-in region (middle two rows). The last two rows show the absolute difference between the ground truth and the reconstructed images.} 
    \label{fig:kernelmerged}
\end{figure*}

Figure \ref{fig:error_dim}.A shows the effect of kernel geometry on deep Koopman autoencoder and GRAPPA reconstructions using the obtained SSIM, PSNR and NRMSE metrics. The proposed approach outperforms GRAPPA across the different kernel configurations for each quantitative metric. It can be observed that the mean value of each metric for the proposed approach remains more stable than GRAPPA when the kernel size is changed in the frequency encoding direction. A representative case showing the reconstruction results for the different kernel geometries is provided in Figure \ref{fig:kernelmerged}. For a kernel size of [5, 2] and [3, 4], the noise amplification is visually higher than the other kernel sizes for GRAPPA in the zoomed-in images and in the absolute differences, while the deep Koopman autoencoder reconstruction remains stable for different kernel sizes.\\

\noindent\textbf{Latent space dimension}\\

Figure \ref{fig:error_dim}.B summarises the quantitative metrics obtained for the proposed approach for the tested range of latent-to-input dimension ratios. Model performance increases slightly as the latent-to-input dimension ratio rises from 0.25 with the metrics peaking at 1.25 before decreasing. Figure \ref{fig:dim_merged} shows the effect of changing the latent-to-input dimension ratio for a representative case. The reconstructions of the proposed model over latent-to-input dimension ratios from  0.8 to 1.2 are visually barely distinguishable. For a latent-to-input dimension ratio of 0.2, 0.4 and 2.0, the reconstructions of the proposed approach present aliasing artefacts (slightly less pronounced than RAKI), as highlighted by the zoomed-in images in Figure \ref{fig:dim_merged}. Additionally, lower values of latent-to-input dimension ratio lead to a significant reduction time in the reconstruction times of the deep Koopman autoencoder (Figure \ref{fig:dim_merged}).


\section{Discussion}

This work proposes a deep-learning based $k$-space interpolation technique for accelerated MRI acquisitions. The main contribution of the study is the development of a deep Koopman autoencoder for scan-specific, nonlinear deep learning estimation of unacquired $k$-space points in parallel imaging reconstruction. This is realised by encoding $k$-space neighbourhoods across all coil channels, applying a linear transformation in the latent space, and then decoding the result, thereby simultaneously reconstructing the original $k$-space source points and interpolating the missing $k$-space targets across all coil channels. The autoencoder part in our approach serves two functions; it captures nonlinear relationships between the $k$-space sources and targets via nonlinear encoding, and constrains this nonlinear mapping by minimising the loss not only on the estimated targets, but also on the reconstructed source points after linear interpolation in the latent space.

The proposed method achieves better NRMSE, PSNR and SSIM metrics than RAKI and GRAPPA, and shows visually less noise amplification, reduced artifact level and sharper images compared to the reference techniques. A clinical reader study confirmed these results that the proposed approach was preferred overall and achieved sharper images with reduced aliasing and better overall image quality. Our experiments show that the technique is relatively robust when provided with limited training data, while the noise amplification increases within the reconstructed images of GRAPPA and artifacts are more prominent in the reconstructions of RAKI as fewer training lines are provided. The effect of the autoencoding was assessed by varying the regularization weight $\lambda$, showing that lowering the autoencoding loss contribution can lead to over-smoothness and the appearance of aliasing artefacts. The reconstruction quality of the proposed approach remained stable and better than GRAPPA across different choices of kernel size. The Koopman autoencoder provided a reconstruction quality that is relatively robust to the latent space dimension; however, heavily reducing the latent-to-input dimension ratio can produce blurring and artifacts. Using an overcomplete autoencoder does not improve the image quality and increases computation time.

Unlike RAKI where the objective function in the training of the network is solely based on the estimation of the missing target $k$-space points, the proposed approach is based on the autoencoding of source $k$-space neighbourhoods and can be understood as a manifold interpolation operation. The autoencoding operation constrains the nonlinear mapping and interpolation in the latent space to minimise the loss after decoding, and acts as an additional consistency constraint, similar to the self-consistency constraint used in SPIRiT \cite{lustig2010spirit} which ensures that the linear shift-invariant property is enforced in the reconstruction across all k-space. This constraint on the nonlinear mapping may explain the improved preservation of details, as well as the reduced level of aliasing artifacts obtained with the deep Koopman autoencoder compared to RAKI. These results concur with the findings of the nonlinear GRAPPA study \cite{Chang2011}, where it was observed that using kernel functions that are too complex can lead to nonlinear GRAPPA reconstructions that suffer from an increased level of blurriness and aliasing artifacts. Similar to other iterative reconstruction techniques, an appropriate regularization weight has to be chosen for the proposed technique, and this parameter can be dependent on the acquisition \cite{lian2025adaptive} (e.g. acquired contrast). As shown in our experiments, if set too low, this parameter can lead to an embedding that is too “lossy” (good denoising but increased blurriness and artifacts), while it could result in insufficient denoising if set too high. Recent approaches based on RAKI \cite{dawoodimage, nimje2025insights} have proposed to tune the level of artifacts and denoising. Dawood et al. \cite{dawoodimage} developed a complex-valued RAKI architecture and used a parameterised leaky $\mathbb{C}$ReLU to control the level of nonlinearity. On the other hand, residual RAKI (rRAKI) and its complex-valued counterpart crRAKI \cite{nimje2025insights} combine two neural network branches (one linear, one nonlinear) that are jointly optimized. The nonlinear network performs nonlinear denoising and aliasing artifact removal from the linear reconstruction residual. The rRAKI architecture may suffer from overfitting when a limited number of ACS lines is available \cite{zhang2022residual, nimje2025insights}. The crRAKI \cite{nimje2025insights} approach alleviates this problem, however with a lower number of autocalibration lines, it foregoes nonlinearity to avoid the enhancement of aliasing artifacts, which can be detrimental for denoising \cite{dawoodimage}. In contrast, our implementation uses a single lightweight Koopman autoencoder architecture providing inductive bias and autoencoding regularization, which can improve robustness to a reduction in the number of ACS lines. One advantage of the proposed method compared to nonlinear GRAPPA \cite{Chang2011, lyu2018kernl, chang2018kernel, wang2019improving} methods is that it is not based on an ad hoc choice of a kernel function or a basis set of particular kernels; instead, the nonlinear mapping in the proposed approach is learned in the encoder and jointly optimized with the interpolation weights.

GRAPPA reconstruction depends on the choice of kernel geometry \cite{Huang2006}. Depending on the smoothness of the coil sensitivity maps, a larger or smaller kernel may yield better reconstruction results\cite{Huang2006}. We have found the deep Koopman autoencoder to be relatively robust to the size of the $k$-space neighbourhoods in our experiments. The nonlinear mapping may help with a kernel size that is too small, reducing truncation-related errors \cite{Chang2011}, while the autoencoding constraint may help with the increase in the size of parameter space when the kernel size is too large.

We have chosen in our implementation to set the dimension of the latent space to 80\% of the dimension of the input space, and to rely on the ModReLU activation functions as a sparsifying operation with biases optimised per node in the neural network. However, our experiments suggest that a level of dimensionality reduction may be achieved to reduce computational complexity of the approach. These results are in line with RAKI \cite{Akakaya2018} and techniques based on low rank properties of $k$-space neighbourhoods \cite{uecker2014espirit, zhang2011parallel, haldar2013low, shin2014calibrationless} and with coil compression techniques \cite{zhang2013coil, chang2018kernel}.

Extensions to RAKI \cite{zhang2022residual} improve its interpretability by performing the RAKI interpolation on the residual of the linear GRAPPA fit. In contrast, the proposed method can be interpreted as a manifold interpolation technique and builds on techniques for modelling nonlinear dynamic systems. Manifold interpolation corresponds to finding a nonlinear mapping where if a linear interpolation is performed and then the result mapped back to the original space, the obtained result is geometrically meaningful, i.e. lies on the manifold. Manifold interpolation has been previously used for interpolating 2D/3D objects and shapes \cite{mayrhofer2023linear} and has found applications in MRI reconstruction, including for k-space to image domain reconstruction (AUTOMAP \cite{zhu2018image}) or to recover dynamic images with a smoothness regularization on manifolds \cite{zou2021dynamic}. The proposed approach uses instead deep Koopman autoencoders to perform nonlinear $k$-space interpolation using $k$-space neighbourhoods as inputs and outputs. Masked autoencoders (MAE) \cite{pan2023global} have also been explored to interpolate Cartesian undersampled $k$-space dynamic data. However, the MAE approach supposes a single coil measurement, and is trained using an external database. The proposed approach is scan-specific and takes into account multi-coil information.

The deep Koopman autoencoder uses $k$-space neighbourhoods that are extracted across all coils, and therefore by design (due to the autoencoding operation), reconstructs the missing targets from all coils. This means that only one network is trained, while RAKI trains two networks per coil channel to reconstruct $k$-space real and imaginary parts retrospectively. However, because RAKI trains a separate network for each coil channel, the reconstruction can be parallelised \cite{zhang2019optimized}, which reduces training and inference times. For both the proposed approach and RAKI, training time impacts the reconstruction time, because both approaches are self-calibrated using autocalibration signals.

This study has several limitations. The reconstruction performance of our approach was only studied with brain scans from the NYU fastMRI \cite{zbontar2018fastmri,knoll2020fastmri} dataset. Further experiments including other anatomies, scanners and coils would be beneficial. The weight of the autoencoding loss was selected manually for each contrast based on visual inspection; automated and adaptive weight selection \cite{lian2025adaptive} could be investigated. We did not investigate the effect of using alternative encoding/decoding architecture (such as multiple layers). A deeper network may learn a more complex nonlinear mapping, however this increases computational complexity, and may lead to overfitting. Self-supervised methods such as SSDU\cite{yaman2020self} with external training datasets or its scan-specific version ZS-SSL \cite{yaman2022zero} produce excellent results with heavier computational complexity and memory footprint.
Moreover, only the ModReLu activation function was evaluated, however other complex-valued activation functions \cite{cole2021analysis, virtue2019complex, daval2019exploring, trabelsi2017deep, virtue2017better, scardapane2018complex} require further evaluation. A further limitation is that the evaluation was performed at a single acceleration factor. In addition, our current reconstructions are performed offline, however for a clinical translation of this work, an inline implementation is required, with optimised reconstruction times. Furthermore, our approach has only been applied for ACS lines acquired using the same pulse sequence as the other $k$-space lines. The proposed approach can also be trained using an external reference scan instead. Another limitation was that training was based on an MSE loss in $k$-space. The learning of low-frequency components may have dominated over the high-frequency content (in the frequency-encoding direction) within the ACS. Several approaches could be investigated to address this: high-dynamic range loss \cite{pan2023global, terpstra2026fast}, nonlinear activation with a compression property \cite{chakraborty2026frequency, scardapane2017kafnets, scardapane2018complex, hammernik2018learning, daval2019exploring}, and cascade training \cite{pan2023global} found in MambaRecon \cite{korkmaz2025mambarecon} and MambaRoll \cite{Kabas2026_ca521be2} or Low-to-High Frequency Progressive learning \cite{xing2024low}.\\

\section{Conclusion}
In this work, we have introduced a scan-specific $k$-space interpolation technique based on deep Koopman autoencoders. Quantitative evaluation based on reconstruction quality metrics and a clinical reader study have shown promising results for the reconstruction of retrospectively accelerated brain MRI datasets.
\FloatBarrier
\bmsection*{Acknowledgments}

This study represents independent research supported by:
\begin{enumerate}
    \item EPSRC Research Council, part of the EPSRC DTP, Grant Ref: EP/W524475/1 and Siemens Healthineers.
    
    \item The National Institute for Health and Care Research (NIHR) Clinical Research Facility (CRF) and HealthTech Research Centre in Cardiovascular and Respiratory Medicine (HRC) at Guy’s and St Thomas’ NHS Foundation Trust. The views expressed are those of the author(s) and not necessarily those of the NHS, the NIHR or the Department of Health and Social Care.

\end{enumerate}

\bmsection*{Conflict of interest}
The first author is a PhD student, and their PhD studentship is co-funded by Siemens Healthineers.
Dr Sarah McElroy, Dr Raphael Tomi-Tricot and Dr Antoine Naegel are employees of Siemens Healthineers.

\bibliography{bibliography}

\clearpage
\onecolumn
\section{Supplementary Materials}
\setcounter{figure}{0}
\renewcommand{\thefigure}{S\arabic{figure}}
\begin{figure}[!htbp]
    \centering
    \includegraphics[width=0.45\textwidth]{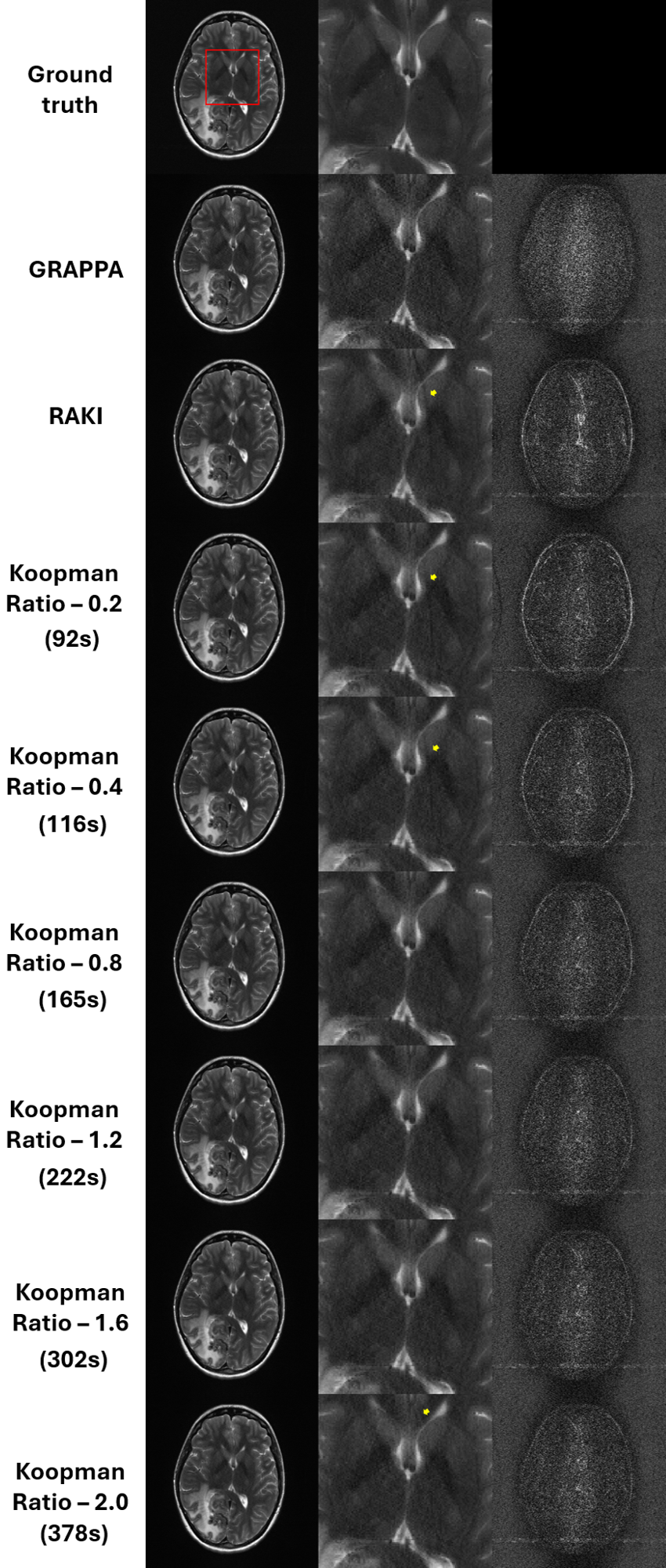}
    \caption[Deep Koopman autoencoder reconstruction using different latent-to-input dimension ratios]{
    Deep Koopman autoencoder reconstruction using different latent-to-input dimension ratios, compared to ground truth, GRAPPA and RAKI. For each method, a reconstruction of a brain image is shown (first column) as well a zoomed-in of the reconstruction region highlighted with a red square on the ground truth image (second column). The third column shows the absolute difference between the ground truth and the reconstructed images. Reconstruction times are provided between brackets for each latent-to-input ratio. Yellow arrows highlight aliasing artifacts on the reconstructed images.} 
    \label{fig:dim_merged} 
\end{figure}

\end{document}